# Practical Multiwriter Lock-Free Queues
# for "Hard Real-Time" Systems without CAS

*Jeremy Lee* BCompSci (Hons)
*The Unorthodox Engineers*

September 9, 2007

## Abstract

FIFO queues with a single reader and writer can be insufficient for "hard real-time" systems where interrupt handlers require wait-free guarantees when writing to message queues. We present an algorithm which elegantly and practically solves this problem on small processors that are often found in embedded systems. The algorithm does not require special CPU instructions (such as atomic CAS), and therefore is more robust than many existing methods that suffer the ABA problem associated with swing pointers. The algorithm gives "first-in, *almost* first-out" guarantees under pathological interrupt conditions, which manifests as arbitrary "shoving" among nearly-simultaneous arrivals at the end of the queue.

## Introduction

There exist several "wait-free" algorithms for common data structures like singly-linked lists, but they generally rely on atomic CAS (compare-and-swap) instructions provided by the processor hardware. Also, they often require "hidden nodes" to be created during queue operations which may have consequences if the memory allocator is not wait-free as well.

Hard Real-Time systems have a requirement to be able to add existing objects to queues during interrupts regardless of the existing queue state. (To be "Re-Entrant to Interrupts") Examples might be a system log for interrupt errors, or a memory allocator command queue. This initially looks like a requirement for "wait-free" code.

A literal interpretation of Herlihy's paper suggests that a generic wait-free multiwriter queues are impossible without hardware support for atomic updates. But "wait-free" is a strong term which assumes robust behavior in the presence of multiple threads running on multiple CPU's at different rates. On a single CPU system, many of these subtleties do not arise.

Single-CPU systems have for years implemented generic multireader/writer queues (in apparent violation of Herlihy) by using the ability to disable interrupts temporarily to avoid pathological conditions. However, this takes them out of the class of real-time systems, since there exist periods, however brief, where the time to process a hardware interrupt will vary depending on the operation already in progress.

Consider a device which must sample an input at a fixed interval (set by a timer) or when a hardware control signal arrives. There is inevitable delay between when the event really occurs and when the processor performs consequent operations, but if interrupts are disabled to allow atomic operations, then this delay can become variable, which introduces jitter into the sampling pattern. Or worse, interrupts could take arbitrarily long to be re-enabled, causing skipped events.

To be a hard real-time system, interrupts must never be disabled. Fortunately, interrupts are a special form of multi-tasking, in that they (usually) run to completion entirely before control is returned to the interrupted task. This is a crucial difference from the more generic multitasking abstractions, and is exploited ruthlessly.

We have created a algorithm which satisfies all these constraints, and although simple, seems to be novel. It is possible it has been implemented out of necessity before, but no published papers that describe a matching algorithm could be found, and a survey of the field suggests that this is an outstanding problem, or at least the solution is not well known.

The presented algorithm is only "wait-free" for interrupts, which provide a crucial guarantee that higher-level interrupts always complete before lower-level interrupts resume. It is not intended for use in a multi-CPU system, or across multiple pre-emptive tasks. This technically weakens the "wait-free" property to "lock-free", but as a practical matter task-switching can easily be delayed during critical periods, whereas interrupts can not.

Furthermore, the algorithm only requires one extra "sentinel" node per queue (which can be the queue itself in a sensible implementation) and a fixed single pointer per node (as with any linked list) to implement the chain, so compatible objects can be passed through multiple queues with no memory allocator activity. This can be crucial for systems like a memory manager command queue, where you don't want the act of managing memory to recursively cause new memory allocations.

The algorithm was designed to run on the dsPIC microcontroller, which does not have an atomic CAS (compare-and-swap) instruction that can work on pointers, which is otherwise common in modern processors. (although the dsPIC has limited single-bit variants) This prevents the use of swing-pointers used in many wait-free algorithms, but interestingly also prevents the ABA problem from occurring, (which is a known but rarely encountered bug in systems based on CAS) thus making the presented algorithm more robust.

While the presented queues are wait-free for all operations and never loose entries, they can suffer from "stalls" where some enqueues by higher-level interrupts will not always be available for dequeue until a lower-level interrupt is continued. This means the queue does not guarantee exact event order in all circumstances, and in fact the algorithm explicitly puts later nodes into the queue before older nodes during special conditions that can occur at high interrupt rates with multiple priorities.

Note that this algorithm requires a synchronized dequeue, which means an assumed single reader thread and/or locking system to prevent read contention. This is rather the opposite of things like shared memory, where you generally synchronize the writer while allowing asynchronous reads.

Since these queues are intended for interrupt-safe system tasks such as thread scheduling and memory management in a hard real-time system, they are driven by these fairly specialist requirements to fit an environment where a small global table and no extra per-node overhead is preferred at the cost of extra processing time, especially if that time is predictable and bounded.

While the algorithm was developed because none of the existing solutions were quite right for our practical requirements, we employ several ideas from older papers, such as Head and Tail separation and sentinel nodes, as proposed by the venerable Valois.

## Pseudocode

The pseudocode is written in stylized object-oriented C (without semicolons or the ambiguous "=" operator) as functional methods of a queue object, which makes local properties of the queue (such as *this*, *head* and *tail*) seem like globals.

The the two basic queue methods are called (for historical reasons due to Valois) **V** and **P**, and the utility functions are mostly to keep the code short. Where named functions are left undefined, (like *get-interrupt-level*) they are machine-specific. One thing to note is that the **P** operator intrinsically makes use of the **V** operator, but not vice-versa.

The algorithm makes use of a global interrupt level table which maintains a queue and node entry for each possible interrupt level. This allows higher-level interrupts to know about the progress of lower-level enqueues. For a small processor like the dsPIC, this table only requires 16 entries for all hardware possibilities, and most real interrupts are actually confined to the first eight levels.

Inline comments are used to make the code more readable, and to assist in translations.

```
// enqueue, put, write
Queue::V(node) {
        // set the global level entries
        level := get-interrupt-level()
        global_level_node[level] := node
        global_level_queue[level] := this
        // interrupts will now append to our level. We can examine and correct the table in peace.
        prev_level := previous_interrupt_level(level)
        while( prev_level != -1 ) {
                // is the previous level node the tail?
                prev := global_level_node[prev_level]
                if( tail == prev ) {
                        // clear the queue entry from the table (but not the node)
                        global_level_queue[prev_level] := null
                        // go around again with the next level down
                        prev_level := previous_interrupt_level(prev_level)
                } else break
        }
        // are we left with the simple case of no previous active levels?
        if( prev_level == -1 ) {
                // catch the queue's tail, and link ourselves to the end
                follow(tail).next := node
                // update the queue tail, now that we can catch ours
                last := follow(node)
                global_level_node[level] := last
                tail := last
                // higher level interrupts may now clear our level
        } else {
                // has the previous level linked itself into a chain yet?
                anchor := find_anchor(prev_level, prev)
                if( anchor == null ) {
                        // stalled. replace the previous levels' chain
                        chain := prev.next
                        prev.next := node
                        // catch our tail and link to the old chain
                        follow(node).next := chain
                        // the lower continuance will catch our appended nodes
                } else {
                        // anchored. replace the existing chain following the anchor node
                        chain := anchor.next
                        anchor.next := node
                        // we are now reachable. catch our own tail, and link the chain
                        last := follow(node)
                        last.next := chain
                        // can we pre-update the queue tail? (optional, but useful)
                        if( anchor == tail ) {
                                global_level_node[level] := last
                                tail := last
                        }
                }
        }
        // clear the global level entries, during which the tail may move past us
        global_level_queue[level] := null
        global_level_node[level] := null
}
```

```
// dequeue, get, read
Queue::P() {
    while(true) {
        // we can't remove the last queue entry
        if( head == tail ) return null
        if( head.next == null ) return null
        // if the queue head is the sentinel
        if( head == sentinel ) {
            // then move to the next node in the chain
            head := head.next
            // enqueue the sentinel again
            sentinel.next := null
            V(sentinel)
            // and loop around for the next node
        } else {
            // take the head node out of the queue
            node := head
            head := node.next
            node.next := null
            // return the dequeued node
            return node
        }
    }
}

// follow a chain to it's last node, which may just be itself
Queue::follow(chain) {
    while( chain.next != null ) chain := chain.next
    return chain
}

// are there any lower interrupt levels that were also working on the same queue?
Queue::previous_interrupt_level(level) {
    while( --level >= 0 ) {
        if( global_level_queue[level] == this ) return level
    }
    return -1
}

// which node (reachable from the global chains or queue tail) references a given node?
Queue::find_anchor(level, node) {
    while( true ) {
        // find the next level down
        level := previous_interrupt_level(level)
        // which chain represents this level?
        chain := ( level < 0 ) ? tail : global_level_node[level]
        // go through the chain and look for the node
        while( chain != null ) {
            if(chain.next == node) return chain
            chain := chain.next
        }
        // return if we ran out of levels to check, otherwise loop
        if( level < 0 ) return null
    }
}
```

# Algorithm Details

Queues that are never empty have less special cases, so a sentinel node is used to simplify things. Queues start with a head and a tail which point to the sentinel node, and the sentinel.next pointer is initially null.

    **H** := **T** := sentinel

From this simplest of states, **P** operations return null, and cause no queue activity. **V** operations will (eventually, after stalls) cause at least one node to be appended to the queue.

| | |
|---|---|
| **H** := **T**:= sentinel | |
| **H** := sentinel - **T** := nodeA | $V_1$ := nodeA |
| **H** := sentinel - nodeA - **T** := nodeB | $V_2$ := nodeB |
| **H** := sentinel - nodeA - nodeB - **T** := nodeC | |
| | $V_3$ := nodeC |

Subsequent **P** operations will take nodes from the head of the queue, and will invisibly "recycle" the sentinel when encountered.

| | |
|---|---|
| **H** := sentinel - nodeA - nodeB - **T** := nodeC | |
| **H** := nodeA - nodeB - nodeC - **T** := sentinel | $P_1$ := nodeA |
| **H** := nodeB - nodeC - **T** := sentinel | $P_2$ := nodeB |
| **H** := nodeC - **T** := sentinel | $P_3$ := nodeC |

Until there is only the sentinel left

    **H** := **T** := sentinel

These two operations are described in detail below, including their intermediate states, and what happens when the ideal sequence of events is interrupted.

The **V** (Enqueue, Write, Put) operator would ideally start by assigning the new node to the Tail's next pointer, and then updating the Tail pointer to the new node. Since this ideal sequence of events can be interrupted at the worst of times, we keep a table of what each interrupt level was working on. We therefore need to check the table levels below our own priority for a 'hanging' operation. (So lower priority interrupts paradoxically take less time to process on average since they need to check on less cases.)

| | |
|---|---|
| **H** := nodeA - **T** := sentinel | $V_1$ := nodeB |
| **H** := nodeA - **T** := sentinel - $V_1$ := nodeB | |
| **H** := nodeA - sentinel - **T** := $V_1$ := nodeB | |
| **H** := nodeA - sentinel - **T** := nodeB | |

This table search is nearly linear in time with the number of possible interrupt levels, so small microcontrollers with only a few interrupt levels are quite efficient, whereas larger processors such as i686 machines can have hundreds of interrupt levels.

If no other operations were in progress on the same queue, then the ideal sequence of operations is done, with the addition that we 'chase the tails' of the queue tail and the new node during certain times to allow for potential inserts from other interrupts. (We leave an expected space for interrupts to place their nodes, like a 'virtual cursor', and only read from those chains once we provide a better place for the cursor.)

| | |
|---|---|
| **H** := nodeA - **T** := sentinel | $V_1$ := nodeB |
| **H** := nodeA - **T** := sentinel - $V_1$ := nodeB | |
| **H** := nodeA - sentinel - **T** := nodeB | $V_2$ := nodeC |
| **H** := nodeA - **T** := sentinel - nodeB - $V_2$ := nodeC | |
| **H** := nodeA - **T** := sentinel - nodeB - nodeC | |
| **H** := nodeA - sentinel - nodeB - **T** := nodeC | |

Otherwise, the state of the interrupted operation is analysed by looking at it's effects on the queue. A decision is made to either append the new node at the end of the lower level's chain, or to insert it just after the existing tail. If an insert is done the tail is updated, so that the tail progresses at the earliest opportunity.

The order of events is chosen carefully so that there is always somewhere that higher level interrupt entries can suddenly appear without compromising the algorithm, and the tail is also treated as quite volatile.

As each new node is added to the level table, it 'freezes' the contents of any chains that may have accumulated in the table below that level, by forcing higher level interrupts to append to this level's chain while this level reads and modifies those lower-level nodes.

Once we have managed to insert ourselves somewhere after the tail, higher-level interrupts switch to a strategy of inserting before this level (sometimes even pre-updating the Tail) because they know we will now be back to reading our own level's chain to find the last entry that may have been put there by previous interrupts.

    **H** := nodeA - **T** := sentinel                                                                                  **V**$_1$ := nodeB , **V**$_2$ := nodeC
    **H** := nodeA - **T** := sentinel - **V**$_2$ := nodeC
    **H** := nodeA - **T** := sentinel - **V**$_2$ := nodeC - **V**$_1$ := nodeB
    **H** := nodeA - sentinel - **T** := **V**$_2$ := nodeC - **V**$_1$ := nodeB
    **H** := nodeA - sentinel - **T** := nodeC - **V**$_1$ := nodeB
    **H** := nodeA - sentinel - nodeC - **T** := **V**$_1$ := nodeB
    **H** := nodeA - sentinel - nodeC - **T** := nodeB

Such near-simultaneous adds are when nodes can become apparently "re-ordered" compared with the ideal sequence of events.

The **P** (Dequeue, Read, Get) operator is rather simpler, and returns null immediately if the queue is detected to have only one entry according to either definition of having it's Head entry's next pointer equal null, or by having the Head and Tail pointers equal. Both checks are done because asynchronous writes might be in progress, leaving the queue in this apparently inconsistent state.

This means that a queue with only one definite entry seems to be empty, and this reflects the fact that the sentinel node should always be in the list somewhere, so long as everything is consistent. If it's not there, then 'hanging' operations at a lower priority level are still reading that node and would be upset if it suddenly went away.

If the queue has at least two entries, and the first entry is the sentinel node, then the sentinel is dequeued, and then enqueued again. If there are no lower-level interrupts that have been left hanging at a certain critical stage, then the sentinel will be available again immediately (amongst other possible entries from higher-level interrupts). But if any of the lower chains is "stalled" then the sentinel may not be reachable, but will return shortly with at least one other node. These two conditions both cause the procedure to return null, but are not otherwise distinguished. (Although it's possible that an intelligent client may want to know that a new node is in the pipeline, but not yet available.)

This process of re-queueing the sentinel node once it is not the only node is called "sentinel recycling" and is a key idea of the algorithm.

If the queue head is not the sentinel, then the head node is dequeued as you would ideally, by setting the head to it's next node, and then clearing the pointer in the dequeued node for safety's sake. The node is completely out of the queue at this point and can be deallocated or re-used with no consequence to the rest of the algorithm.

This is different from some previous solutions, which also enforce a 'one-node minimum' on their queues by seeding them with a temporary 'first' sentinel, and which return the successor nodes early. This works well if your objects are garbage collected or reference counted, but not so well if you're using the queue to feed, for example, the memory deallocator. This design choice means once the entries are dequeued, they are guaranteed to be no longer referenced by hanging queue operations at any level. Nodes are completely enqueued before they can be dequeued.

### Queue States
The algorithm is designed to keep the queue in a small set of recognised states at all times.

During **V** operations, four states are distinguished via the level table and the queue Tail:

1. There are no lower level nodes in the level table, indicating a consistent queue Tail and unrestricted access to it.

2. The lower level node is not reachable from the Tail or anywhere in the table, indicating that the lower level is still seeking a chain to modify, or the link hasn't finished.
3. The lower level node is reachable from an 'anchor' in a chain, indicating that the lower enqueue has updated it's anchor node's pointer, and is now searching it's own chain.
4. The lower level node is the Tail, indicating the lower enqueue is mostly complete, and was just about to clear the level table entries. (in which case the level is cleared, and the check repeats)

During **P** operations, three states are distinguished by examining the queue Head and Tail:

1. If the Head and the Tail are the same node, or the Head node's next pointer is null then there is only one (unavailable) entry in the queue. (If the Head is not the sentinel, then the queue is 'stalled')
2. If the queue has more than one entry and the first entry is not the sentinel, then it can be dequeued
3. If the queue has more than one entry and the first entry is the sentinel, then the sentinel should be silently recycled.

There are some cases which should never be created by the algorithms, and are not tested for:

1. A queue which consists only of a sentinel node which points to itself. The enqueue operation will infinite loop while trying to 'chase the tail'.
2. A queue which consists of any loops, create by enqueueing a node or chain which links to an existing queue node. V will go into an infinite loop, unless P operations disassemble the loopy queue back into a sane state in time.
3. A queue without a sentinel node, but not in the process of recycling the sentinel. Such a 'lost sentinel' would not be fatal, but would stall one entry on the queue to take the sentinel's place, and is hard to reliably detect.
4. A queue with no nodes at all (Head and/or Tail set to null) should only occur while the queue object is in the process of being created or destroyed. Operations on a partially initialized queue will dereference a null pointer very quickly.
5. A global table level with a queue entry, but no node entry.

Detecting any of these symptoms means the queue is in a fatal state, and unrecoverable without special intervention. Real systems are encouraged to do consistency checks as part of debug or logging methods.

## Outline of Proof

Several earlier papers [Valois] show that a FIFO queue using Head and Tail pointers (and a sentinel node) have V and P operations that are asynchronous with each other. (but not themselves) Leading to implementations like ring-buffers which work very well in practice.

In general terms, if task A only reads and writes the Head, and task B only reads and writes the Tail, then task A and B are asynchronous and wait-free, by the trivial proof that they share no state in common.

In this queue implementation the V operator modifies (reads then writes) the Tail pointer but never accesses the Head pointer. The P operator modifies the Head and only reads the Tail. This makes V wait-free for all states of the P operator.

V updates the Tail pointer always in the direction along the chain, so that reads at time T+1 should always return the same node, as returned at T, or one chained from it. So long as this property holds, the queue Tail will be consistent in time.

The P operator moves the Head pointer along the chain, but never past the Tail. This guards it against accessing nodes which may be in flux. It performs a single read of the Tail pointer to check this condition, creating a weak asynchronous link with the V operator. It may also call the V operator once, creating a strong asynchronous link with V, just to be sure. So long as V is asynchronous and wait-free with itself, then P will be asynchronous and wait-free with V.

Since P only updates the Head pointer to the very next node (sometimes twice) reads at time T+1 will always return a node that was reachable at time T. So long as this property holds, the queue Head will be consistent in time.

The two time consistency guarantees mostly depend on whether the computer hardware re-orders memory writes because of cache, or whether multi-CPU systems lack accurate knowledge of shared memory because of lag. In a single-CPU system that does not re-order memory writes, this time consistency necessarily holds and the queue becomes robust.

## Implementation

The algorithm was implemented in C for Microchip's dsPIC30F series of microcontrollers (using the GNU C compiler provided by Microchip in their MPLab IDE) as a "polymorphic" C function which can operate on a variety of structures using pointer offsets. Functions were renamed slightly to fit an existing scheme, but otherwise are direct translations of the pseudocode into C. (using macros to reference the structure fields)

The compiled size of each function is given in dsPIC instructions.

| | |
|---|---|
| halChainFollow | 11 |
| halChainPreviousLevel | 13 |
| halChainAnchor | 35 |
| halChainEnque | 101 |
| halChainDeque | 37 |

Total size for the library was 201 instructions. Each dsPIC instruction is three bytes long. 603 bytes is a good library size for embedded systems, where the whole device may only have 16K of storage.

## Optimizations

If the Queue object is also a Node, then you can use the queue as it's own cycling sentinel and save on allocating a new memory object per queue initialization.

Auxiliary functions can be inlined into the main methods for speed. They are defined separately only to keep the code readable.

Most attempts to remove the sentinel node from the algorithm just result in an extra node pointer being needed on the queue, which is basically equivalent.

Since the P operator only ever calls V to re-queue the sentinel, (which may be a specialised node type in some implementations) an optimized routine just for this operation might be effective.

A "Peek(n)" operator that checks for the existence of nodes (other than the sentinel) on the queue would be wait-free and asynchronous as long as it did not re-order the queue during the test. (it would just have to skip over the sentinel, rather than recycling it) This could be a useful wait-free precursor to the synchronous P operator, and would prevent waits on empty queues. A "Count" operator would be very similar.

There is no obvious reason why the currently synchronous P function can't be made asynchronous (re-entrant) like V, but this would probably require a second global table (at least) with similar logic to keep the Head in a consistent state across time, and it may not be possible to keep the minimal overhead guarantees of the existing algorithm. One approach is to add a 'read lock' per node (using bitwise CAS instructions available on the dsPIC, or spinlocks which depend on the atomic nature of increment/decrement instructions and their side-effects on the zero flag) to indicate when each node has been consumed. Finally, making the P procedure re-entrant would definitely require some kind of reference-counting or garbage collection to avoid hidden intermediate nodes, or you necessarily loose the "no longer in use" guarantee of the algorithm's P requests. All in all, probably not worth it.

## Conclusions

The FIFO queue with a single producer/consumer pair is a favorite design choice for hardware device queues, and extending the capabilities of the writer by making the enqueue operation lock-free and re-entrant to interrupts actually simplifies the dequeue operation. Absolute minimal overhead is found on small systems with relatively few interrupt levels, which are also the systems most likely to lack atomic CAS capability.

Such queues, while not technically "First In, First Out" for some definitions of "First In", are very useful when absolute order is not strictly important, such as interrupt-driven notification queues where the notion of "first" can get fuzzy anyway.

Multi-writer queues have an interesting advantage over multi-reader queues, in that write operations can intrinsically provide the storage structures that will be used in the course of the action. Multi-reader queues generally have to create extra 'hidden' objects to support their functions, which adds an extra (unpredictable) cost in time and memory. Multi-writer queues 'amortise' these overheads upfront as part of their node creation cost.

These queues create a very versatile primitive in hard real-time systems, and can be used to construct "N-way Queues" where an intermediate "copier" thread distributes the results of a single multiwriter input queue to N registered reader queues. These are potentially a universal messaging primitive within the limitations of a real computer system.